\documentstyle[aps,prb,epsf]{revtex}
\begin{document}
\draft
\wideabs{
\title{Finite Size Effects on Spin Glass Dynamics}
\author{Y.G. Joh}
\address{National High Magnetic Field Laboratory - Los Alamos National Laboratory, NM 87545}
\author{R. Orbach, G.G. Wood}
\address{Department of Physics, University of California, Riverside, California  92521-0101}
\author{J. Hammann, and E. Vincent}
\address{Service de Physique de l'Etat Condens\'e, Commissariat \`a l'Energie Atomique,
Saclay, 91191 Gif sur Yvette, France}
\date{Submitted February 5, 2000}
\maketitle
\begin{abstract}
The recent identification of a time and temperature dependent spin glass correlation length, $\xi({t_w},T)$, has consequences for samples of finite size.  Qualitative arguments are given on this basis for departures from $t/{t_w}$ scaling for the time decay of the thermoremanent magnetization, ${M_{TRM}}(t,{t_w},T)$, where $t$ is the measurement time after a ``waiting time'' $t_w$, and for the imaginary part of the ac susceptibility, ${\chi^{{\prime}{\prime}}}(\omega,t)$.  Consistency is obtained for a more rapid decay of $M_{TRM}(t,{t_w},T)$ with increasing $t_w$ when plotted as a function of $t/{t_w}$, for the deviation of the characteristic time for ${M_{TRM}}(t,{t_w},T)$ from a linear dependence upon $H^2$ at larger values of $H$, and for the deviation of the decay of ${\chi^{{\prime}{\prime}}}(\omega,t)$ from $\omega t$ scaling upon a change in magnetic field at large values of $\omega t$.  These departures from scaling can, in principle, be used to extract the particle size distribution for a given spin glass sample.
\end{abstract}
}
\narrowtext

\section{Introduction}

Slow spin-glass dynamics have been investigated through examination of the time dependence of the irreversible magnetization,\cite{Lundgren} and the out-of-phase dynamical magnetic susceptibility.\cite{Refregier}  Both are related, but typically explore different time domains.  A recent paper\cite{Wood} has shown how the Parisi order parameter,\cite{Parisi} $x(q)$, can be extracted from high precision measurements of the decay of the thermoremanent magnetization, ${M_{TRM}}(t,{t_w},T)$, where $t$ is the measurement time after waiting a time $t_w$ at the measurement temperature $T(<{T_g})$.  The time scale of the experiments restricts the examination to a small region of overlap space, but the shape and magnitude of the results are in quantitative accord with mean field expressions.\cite{Mezard}

The barrier model of Nemoto\cite{Nemoto} and Vertechi and Virasoro\cite{Vertechi} has been adapted to the metastable states observed in experiment.\cite{Lederman}  Dynamics are ascribed to barrier hopping within ultrametric manifolds of constant magnetization $M$.\cite{Mezard2} Measurements by Chu {\it et al.},\cite{Kenning} along with other observations of a similar nature,\cite{Bouchaud} suggest that a change in magnetic field {\it reduces} the barriers in the initially occupied magnetization manifold.  This reduction can be conceptually thought of as diffusion between states of constant $M$ through intermediate states of lower Zeeman energy.\cite{Bouchaud}  We designate the reduction in each of the barrier heights by a Zeeman energy, $E_z$, proportional to the number of participating spins, $N_s$, which lie within a coherence length $\xi({t_w},T)$ from one another.  This length scale has been extracted from measurements on both insulating and metallic spin glasses.\cite{Wood2}  As shown in Ref. 12, and exhibited below in this paper as a consequence of further measurements, the spin glass correlation length appears {\it universal}, having the value $\xi({t_w},T)=({t_w}/{\tau_0})^{\alpha T/{T_g}}$, with $\alpha=0.153$, for all spin glasses measured so far ($Cu:Mn, Ag:Mn$, and the thiospinel $Cd{Cr_{1.7}}{In_{0.3}}{S_4})$.  The length $\xi({t_w},T)$ is in units of the typical spin-spin spatial separation.

This paper examines the consequences of a finite length scale associated with spin glass order.  In particular, we address the question of what happens when $\xi({t_w},T)$ exceeds a physical length $r$ associated with finite sample size ({\it e.g.} defects in an insulating structure, or crystallites in a polycrystalline sample).  Typical laboratory waiting times and measurement temperatures result in $\xi({t_w},T)\approx 10-100~nm$, approximating the coarseness of powdered or polycrystalline samples.  Such materials do not possess a single grain size $r$, but rather a distribution of grains sizes, $P(r)$.  This paper will show that it may be possible to extract $P(r)$ from time dependent magnetic measurements.

A consequence of $\xi({t_w},T)\approx~r$ is that conventional scaling relationships may be violated.  This paper will associate the lack of scaling with $t/{t_w}$ found for the time decay of ${M_{TRM}}(t,{t_w},T)$ by Alba {\it et al.}\cite{{Alba},{Ocio}} with $\xi({t_w},T)\approx~r$.  Other observations which we believe can be associated with
$\xi({t_w},T)\approx~r$ are the deviation of $E_z$ from a proportionality to $H^2$,\cite{{Wood2},{Vincent2}} and the deviation of ${\chi^{{\prime}{\prime}}}(\omega,t)$ from $\omega t$ scaling upon a change in magnetic field for large $\omega t$.\cite{Bouchaud}

This paper will attempt to make plausible the relationship between finite size and lack of scaling in the same spirit as in Bouchaud {\it et al.}\cite{Hammann}  Sec. II describes the model which underlies our analysis.  Sec. III outlines the experimental observations which require a departure from scaling.  The relationships between finite size effects and the lack of scaling are developed in Sec. IV.  Key experiments and numerical simulations required to fully justify these relationships are discussed in Sec. V.  Sec. VI presents our conclusions.

\section{Description of the Model}

The model we invoke to quantify our analysis is based on an extension of the ``pure states'' ultrametric geometry found\cite{Mezard3} for the mean field solution\cite{Parisi} of the Sherrington-Kirkpatrick infinite range model.\cite{Sherrington}  Previous experiments\cite{Lederman} have shown that the infinite barriers which separate pure states derive from finite barriers separating metastable states, diverging at a characteristic temperature $T^*$ as the temperature is lowered.  Within the time and temperature scale probed experimentally,\cite{Lederman} the value of $T^*$ depends only on the height of the barrier at a given temperature, and a universal form for $d\Delta/dT~vs~T$ was derived from experiment.

Dynamics are extracted from the assumption that the metastable states possess the same ultrametric geometry as pure states, the barriers separating the metastable states increasing linearly as the Hamming Distance $D$ between states,\cite{Nemoto,Vertechi} $\Delta(D)\propto D$, and activated dynamics.

The time decay of the thermoremanent magnetization follows from assuming that a temperature quench in constant magnetic field isolates the spin glass states into specific points within phase space.  This is represented by a probability density $P(D)$, with $D$ the Hamming distance defined by $D={1\over 2}({q_{EA}}-q)$, where $q_{EA}$ is the Edwards-Anderson order parameter,\cite{Edwards} and $q$ the overlap between the states separated by $D$.  Immediately after quench, $P(D)=\delta(D)$.  Keeping the magnetic field constant, the system diffuses from $D=0$ to states with $D\neq 0$ as a function of the ``waiting time'', $t_w$, according to activated dynamics.  As stated earlier, the ultrametric tree upon which this diffusion develops is one of constant magnetization $M_{fc}$ associated with the field cooled state.\cite{Mezard2}  The amplitude of the delta function diminishes with increasing $t_w$, with the associated occupancy of states at finite $D$ increasing with $t_w$.  Activated dynamics in phase space leads to a maximum barrier surmounted in the time $t_w$ of magnitude $\Delta({t_w},T)={k_B}T\ell n({t_w}/{\tau_0})$.  Detailed balance, with experimental confirmation,\cite{Lederman,Kenning} leads to equilibrium occupation of the metastable states.

Experiments and analysis by Vincent {\it et al.}\cite{Bouchaud} establish that, after waiting a time $t_w$, cutting the magnetic field to zero diminishes each barrier height by an amount to which we shall refer as $E_z$, the {\it Zeeman energy}.  This reduction was interpreted\cite{Lederman} as the origin of the so-called ``reversible'' change in the magnetization.  Their model assumes that all of the states occupied for $D<{D_{E_z}}$ [including those at $P(0)$] immediately empty to the zero magnetization manifold $M=0$.  If $\Delta(D)$ is known, then $D_{E_z}$ is the value of $D$ at which ${E_z}=\Delta(D)$.  Field cycling experiments show\cite{Lederman} that $D$ is ``respected,'' that is, the population of the states within $0\leq D<{D_{E_z}}$ in the $M$ manifold rapidly transitions to the states $0\leq D<{D_{E_z}}$ in the $M=0$ manifold.  At the measurement time $t$, where the ``clock'' starts when the magnetic field is cut to zero, {\it i.e.} at $t_w$, the population of the states remaining behind for $D>{D_{E_z}}$ in the $M$ manifold have a total magnetization  ${M_{TRM}}(t,{t_w},T)$.  They decay to the $M=0$ manifold by diffusing from $D>{D_{E_z}}$ to the ``sink'' created by the magnetic field change $D<{D_{E_z}}$.  The characteristic response time is set by the time it takes for the population in the states which have surmounted the highest barrier, $D_{\Delta({t_w},T)}$, to diffuse to the Hamming distance at the edge of the sink, $D_{E_z}$.  By supposition, this is equivalent to surmounting a barrier of characteristic height $\Delta({t_w},T) - {E_z}$.

For very small $E_z$, this diffusion process yields\cite{Nordblad} a peak in the spin glass relaxation rate
$S(t)=d[-{M_{TRM}}(t,{t_w})/H]/d\ell n~t$ at $t\approx{t_w}$.  For finite $E_z$, the peak in $S(t)$ is shifted to shorter times ${t_w^{e\!f\!f}}$, and was first noted for experiments upon the insulating thiospinel $\rm {Cd{Cr_{1.7}}{In_{0.3}}{S_4}}$ by Vincent {\it et al.}\cite{Bouchaud} and the amorphous system $\rm {{({Fe_x}{Ni_{1-x}})_{75}}{P_{16}}{B_6}{Al_3}}$ by Djurberg {\it et al.}.\cite{Djurberg}  The effective characteristic time immediately follows from activated dynamics,
$$\Delta({t_w},T)-{E_z}={k_B}T(\ell n~{t_w^{e\!f\!f}}-\ell n~{\tau_0})~~.\eqno(1)$$

The final piece of the puzzle is the magnitude of $E_z$.  The overall magnetization before the magnetic field is cut to zero, $M_{fc}$, is essentially constant at the measurement temperature during the waiting time $t_w$.  One can think of $M_{fc}$ as arising from the population of the states in the $M$ manifold, each of which has the same magnetization.  If we define the susceptibility {\it per spin} in the field cooled state as $\chi_{fc}$, then the magnetization per occupant of each state is just ${\chi_{fc}}H$.  The Zeeman energy, $E_z$, is the amount by which the barriers in the $M$ manifold are reduced.  Bouchaud\cite{Bouchaud} ascribes this ``reduction'' to diffusion {\it out of the constant $M$ plane}, caused by a ``tilting'' of the overall energy surface.

For the barriers $\Delta(D)$ to be uniformly reduced, there must be a coherence associated with the ``hopping'' process.  That is, there must be a certain number of spins, $N_s$, which are rigidly locked together, and which participate in the hopping process as a coherent whole.  But these spins possess a net magnetization ${N_s}{\chi_{fc}}H$, or a Zeeman energy ${E_z}={N_s}{\chi_{fc}}{H^2}$.

Hence, use of Eq. (1) in experiments as a function of $H$ generates {\it absolute values} for ${E_z}$, and thence for $N_s$, because $\chi_{fc}$ is known.  It was this analysis which enabled Joh {\it et al.},\cite{Wood2} using $N_s\propto{\xi(t,T)^3}$, to find $\xi({t_w},T)$, and to compare with Monte Carlo calculations\cite{Parisi2} by extrapolating to time scales ten orders of magnitude shorter than laboratory times.

Measurements of ${log_{10}}{t_w^{e\!f\!f}}~vs~{H^2}$, from Eq. (1), should yield a straight line, the slope of which can be used to obtain an absolute value for ${N_s}(t,T)$, and thence $\xi(t,T)$.  As will be shown in the next Section, the $H^2$ dependence of ${log_{10}}{t_w^{e\!f\!f}}$ is found only at the small end of the magnetic field range.  At larger magnetic fields, the dependence on $H$ veers away from quadratic to more like linear.  Further, dynamics derived from this model should scale as $\displaystyle{t\over {t_w}}$, whereas the data quoted by Joh {\it et al.},\cite{Wood2} Bouchaud {\it et al.},\cite{Hammann} and Vincent {\it et al.},\cite{Bouchaud} do not.  Sec. III (following) displays the results of experiments which exhibit these deviations, forming the basis for our subsequent analysis in Sec. IV where we account for these deviations on the basis of finite size effects, along the same lines previously proposed by Bouchaud {\it et al}.\cite{Hammann}

\section{Experimental Evidence for Lack of Scaling}

The previous Section showed that Eq. (1) could be used to obtain a quantitative value for the Zeeman energy ${E_z}={N_s}{\chi_{fc}}{H^2}$.  This requires ${E_z}$ to scale as $H^2$.  Fig. 1 reproduces Fig. 2 of Ref. 12.

\begin{figure}
\epsfysize=2.3in \epsfbox{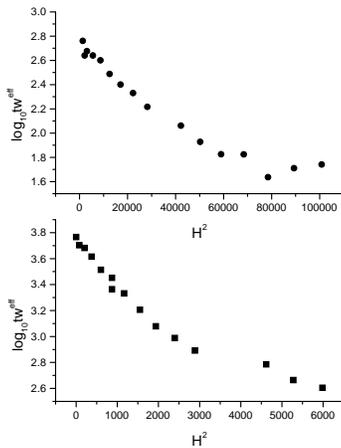}
\caption{A plot of
${log_{10}}{t_w^{e\!f\!f}}$ [equivalently ${E_z}$ from Eq. (1)]
{\it vs} $H^2$ for $Cu:Mn~6~at.\%$ ($T/{T_g}=0.83,~{t_w}=480~{\rm
sec})$ and the thiospinel, $Cd{Cr_{1.7}}{In_{0.3}}{S_4}$,
($T/{T_g}=0.72,~{t_w}=3410~{\rm sec})$ at fixed $t_w$ and $T$ from
Ref. 12.  The dependence is linear in $H^2$ for magnetic fields
less than 250 and 45 $G$, respectively, then ``breaks away'' to a
slower dependence.  The scale for $H^2$ in the two plots has been
adjusted to the distance from the de Almeida-Thouless line at the
respective temperatures.} \label{fig1}
\end{figure}

This figure exhibits a quadrative dependence for ${log_{10}}{t_w^{e\!f\!f}}$ on $H$ in the very small magnetic field change limit, ``breaking away'' to a slower dependence at a slightly larger field, $H_{break}$.  We shall associate $H_{break}$ with the smallest crystallite size.  Alternatively, a linear dependence on $H$ can be fitted to the data on the thiospinel\cite{{Bouchaud},{Vincent2}} at over a range from small to moderate $H$, with a deviation at very small $H$.

Taking the small $H$ region slope from Fig. 1, together with additional data recently obtained by the authors (thiospinel, measured at UC Riverside; and $Ag:Mn 2.6~at.\%$, measured at SACLAY), allows one to plot $N_s~vs~(T/{T_g})\ell n({t_w}/{\tau_0})$ for three different physical systems, over a wide range of reduced temperatures, $T_r$, and waiting times $t_w$.  The results are exhibited in Fig. 2.  The solid line drawn through the points, setting $\xi({t_w},T)={{N_s}^{1\over 3}}$, is the relationship quoted in the Introduction,

$$\xi({t_w},T)=({t_w}/{\tau_0})^{0.153T/{T_g}}~~,\eqno(2)$$
where the unit of length is the typical spin-spin spatial separation.
\begin{figure}
\vspace{1.0cm} \epsfysize=2.6in \epsfbox{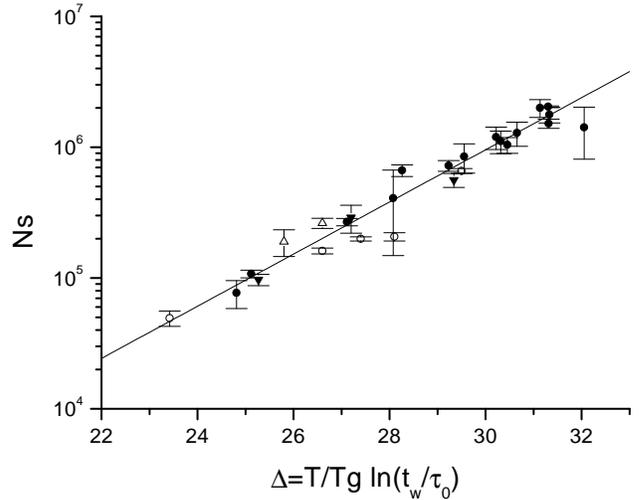} \caption{A plot
of $N_s$ on a log scale {\it vs} $(T/{T_g})ln({t_w}/{\tau_0})$ for
$Cu:Mn~6~at.\%$ (solid circles); thiospinel measured at SACLAY
(open triangles); thiospinel measured at UC Riverside (solid
inverted triangles); and $Ag:Mn 2.6~at.\%$ (open circles).}
\label{fig2}
\end{figure}

In a similar fashion, the decay of the thermoremanent magnetization ${M_{TRM}}(t,{t_w},T)$ does not obey simple $\displaystyle{t\over {t_w}}$ scaling.  Fig. 3 reproduces Fig. 1b of Ref. 23, a plot of ${\displaystyle{{M_{TRM}}(t,{t_w},T)\over {M_{fc}}}}~vs~{\displaystyle{t\over {t_w}}}$, but with an expanded scale.  The failure to scale with $t/{t_w}$ is seen clearly.

The lack of scaling arises from two sources.  The first is emptying of the delta function at $D=0$ with increasing $t_w$, appropriate to the barrier model; or, equivalently, from the presence of stationary dynamics in the model of Vincent {\it et al.}\cite{Vincent3}.  The second contribution arises from transitions between barriers, or, equivalently from the non-stationary dynamics.

In order to display the part of the decay of $\displaystyle{M_{TRM}(t,{t_w},T)\over {M_{fc}}}$ associated with the dynamics of barrier hopping, or, concomitantly, the non-stationary part of the magnetization decay, the data of Fig. 3 are replotted in Fig. 4 with the estimated stationary contribution of Vincent {\it et al.}\cite{Vincent3} subtracted from the full measured value.

The behavior exhibited in Figs. 3 and 4 was ascribed to an ``ergodic'' time, $t_{erg}$ by Bouchaud {\it et al.}\cite{{Bouchaud},{Hammann}}, associated with the occupation of the deepest trap  in a ``grain'' with finite numbers of states.  We shall argue that a finite $t_{erg}$ is also responsible for the behavior exhibited in Fig. 1.

\begin{figure}
\epsfysize=2.5in \epsfbox{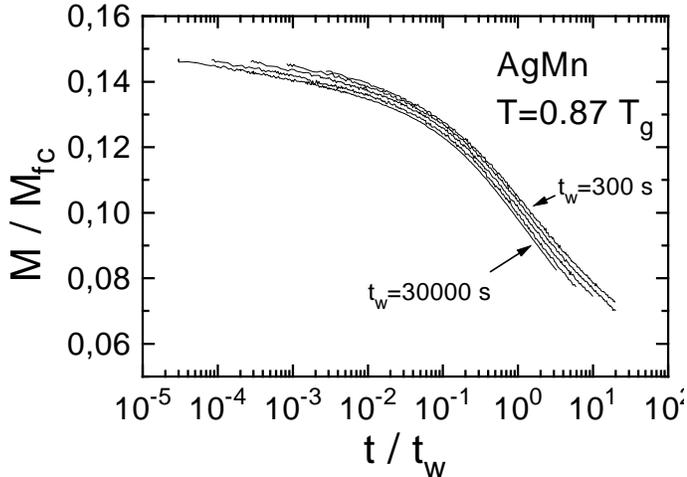}
\caption{The decay of
$\displaystyle{{M_{TRM}}(t,{t_w},T)\over {M_{fc}}}$ as a function
of ${log_{10}}(t/{t_w})$ for $Ag:Mn~2.6~at.\%$ for ${t_w}=$ 300,
1000, 3000, 10000, and 30000 seconds from Refs. 14 and 23. The
lack of scaling with $t/{t_w}$ is evident at large $t/{t_w}$.}
\label{fig3}
\end{figure}

\begin{figure}
\epsfysize=2.3in \epsfbox{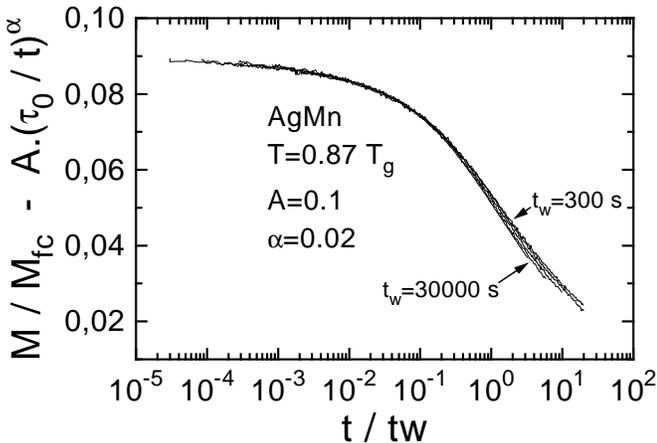} \caption{Same as Fig. 3, but
with the estimated stationary contribution subtracted from the
full measured value. The lack of scaling with $t/{t_w}$ continues
to be evident at large $t/{t_w}$, and it is clearer that
$\displaystyle{{M_{TRM}}(t,{t_w},T)\over {M_{fc}}}$ for longer
waiting times decays faster than for shorter waiting times.}
\label{fig4}
\end{figure}
In addition to the lack of $t/{t_w}$ scaling for the ${M_{TRM}}(t,{t_w},T)$, there is also a departure from scaling for the time dependent magnetic susceptibility.\cite{Bouchaud}

Reproducing Fig. 2 of Ref. 11 in Fig. 5 for the thiospinel, it is seen that there is an $\omega$ dependence for the magnitude of the jump in ${\chi^{\prime\prime}}(\omega,t)$ when plotted as a function of $\omega t$, for $\omega t$ large (of the order of 1,000 or larger), in violation of scaling.  However, for small $\omega t$, scaling is obeyed.  These features will be shown to be consistent with a distribution of crystallites of finite size in the next Section.

\begin{figure}
\vspace{1.5cm} \epsfysize=2.4in \epsfbox{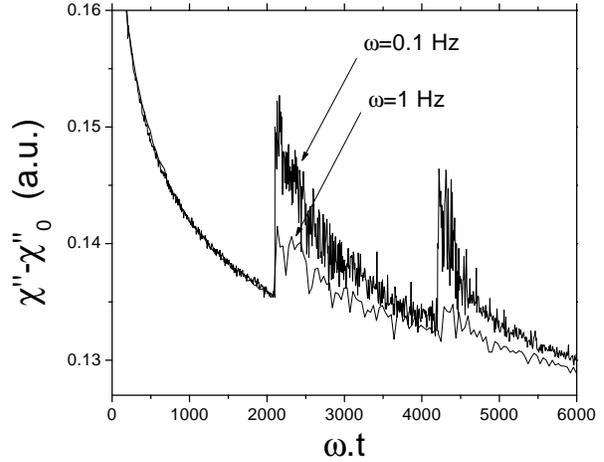}
 \caption{Comparison of the
effect on $\chi^{\prime\prime}$ of a dc field variation in two
experiments at frequencies $\omega=0.1~{\rm and}~1~Hz$ from
Vincent {\it et al.} in Ref. 11. The field variation is applied at
a time $t_1$ (= 350 and 35 $min$, respectively) such that the
product $\omega \cdot {t_1}$ is kept constant.  The curves are
plotted as a function of $\omega\cdot t$, and have been vertically
shifted in order to superpose both relaxations before the field
variation.  At constant $\omega \cdot t$, the effect of the
perturbation is seen to be stronger for the lowest frequency
(longest $t_1$).} \label{fig5}
\end{figure}

In summary, three phenomena show departures from the predictions of the model outlined in Sec. II: (a) the ``break away'' from the $H^2$ dependence of ${log_{10}}{t_w^{e\!f\!f}}$; (b) the lack of scaling with $t/{t_w}$ of the time dependence of the thermoremanent magnetization ${M_{TRM}}(t,{t_w},T)$; and (c) the lack of scaling with $\omega t$ of ${\chi^{\prime\prime}}(\omega,t)$ at large $\omega t$.  These separate, but related, observations are consistent with a particle size distribution $P(r)$ in the sample, such that the spin glass correlation length $\xi({t_w},T)$ becomes comparable with the size of a component particle $r$.

\section{Finite size effects and the lack of scaling}

We examine in this Section the dynamics of a small spin glass particle of radius $r$, at a waiting time and temperature where $\xi({t_w},T)$ is comparable to $r$.  Aging ceases in the particle when this occurs, first noted by Bouchaud et al.\cite{{Bouchaud},{Hammann}}

The barrier model of Sec. II has the following consequences.  For increasing $t_w$, occupied states are separated by barriers which increase in height according to,\cite{Wood2}
$$\Delta({t_w},T)=6.04{k_B}{T_g}\ell n~\xi({t_w},T)~~.\eqno(3)$$
Immediately after the time $t_w$ when the magnetic field is cut to zero, Eq. (3) specifies the {\it maximum} barrier height surmounted by the system.  Should $\xi({t_w},T)\approx r$, there would be no more barriers to surmount!  The occupation of all states would be at equilibrium, and aging in the sense of the barrier model ceases.  Of course, any physical system will have a distribution of particle sizes, $P(r)$, so that this cessation will be ``smeared'' in time.  The purpose of this Section is to explore the impact of $\xi({t_w},T)\approx r$ for each of the three experimental departures from scaling presented in Sec. III.\hfill\break
\par
\noindent
{\bf (a) ``Break away'' from $H^2$ dependence of ${log_{10}}{t_w^{e\!f\!f}}$}\hfill\break
\par
As introduced in Sec. II, a susceptibility, $\chi_{fc}$, can be associated with each spin in the field cooled state, resulting in ${E_z}={N_s}{\chi_{fc}}H^2$.  In this way, the slope of the plot of $E_z$ vs $H^2$ generates $N_s$, the number of spins locked together in a coherent state.  As shown in Sec. III, the actual data do appear to scale as $H^2$ for very small magnetic field changes, but this scaling breaks down at slightly larger fields, ${H_{break}}\simeq 170~G$ for $Cu:Mn~6at.\%$ and ${H_{break}}\simeq 45~G$ for the thiospinel, $Cd{Cr_{1.7}}{In_{0.3}}{S_4}$, with each $H_{break}$ equivalent to about 15\% of the respective de Almeida - Thouless critical field.\cite{Almeida}  An alternative linear dependence of $E_z$ on $H$ does describe the data over the field range beginning with $H_{break}$ and extending to the largest magnetic field change,\cite{Wood2,Vincent2} but fails at very small field change.  At the time of the publication of Ref. 12, we wrote that ``We do not have a satisfactory explanation for this change in slope.''

We believe that the departure of the plot of $E_z$ from proportionality to $H^2$ can be understood within the barrier model of Sec. II, modified to include finite size effects.

Consider a spin glass particle of radius $r$.  The number of correlated spins would be proportional to $r^3$ should $r<\xi({t_w},T)$, but proportional to ${\xi^3}({t_w},T)$ should $r>\xi({t_w},T)$.  Thus, $E_z$ would be less for the smaller particles ($r<\xi({t_w},T)$) than for the larger particles ($r>\xi({t_w},T)$) at the same value of $H^2$.  Further, the largest barrier overcome on a time scale $t_w$ would be $\Delta(r)=6.04{k_B}{T_g}\ell n\,r$ for $r<\xi({t_w},T)$ and $\Delta({t_w},T)=6.04{k_B}{T_g}\ell n\,\xi({t_w},T)$ for $r>\xi({t_w},T)$.  The effective waiting time, $t_w^{e\!f\!f}$, depends upon the difference $\Delta - {E_z}$ from Eq. (1).  For small magnetic field changes, this means that $t_w^{e\!f\!f}$ is larger for the larger particles and smaller for the smaller particles [$\ell n\,\xi({t_w},T)>\ell n\,r$].

As the magnetic field change increases, $E_z$ increases more rapidly for the infinite [meaning $\xi({t_w},T)<r$] size particles than for the smaller [meaning $\xi({t_w},T)>r$] size particles.  This means that the peak in $S(t)=d\Bigl[\displaystyle{{-M_{TRM}}(t,{t_w},T)\over H}\Bigr]\Bigl/d\ell n\,t$ shifts to shorter times {\it more rapidly} with $H$ for the infinite particles than for the smaller particles.  At some value of magnetic field change $H$, $t_w^{e\!f\!f}$ for the infinite and smaller particles will become equal.  This yields an increase in apparent width for $S(t)$,  as observed in many experiments.\cite{Wood2}  For yet larger $H$, the weight of all the smaller particles dominates, slowing the shift of the peak of $S(t)$ with increasing $H$, leading to a less rapid decrease of ${Log_{10}}{t_w^{eff}}$ with increasing $H$.

We believe this to be the origin of the ``break'' in the slope of ${Log_{10}}{t_w^{eff}}$ versus $H^2$, exhibited in Fig. 1 of Sec. III, and therefore an effect of finite size.  A quantitative fit will require knowledge of the particle size distribution $P(r)$.  For now, this finite size effect can explain the behavior of the magnetic field dependence of the characteristic response time\cite{Bouchaud,Wood2,Vincent2}
within the barrier model of Sec. II.\hfill\break
\par
\noindent
{\bf (b) Lack of $\displaystyle {t\over {t_w}}$ scaling for ${M_{TRM}}(t,{t_w},T)$}\hfill\break
\par
The barrier model of Sec. II to describe spin glass dynamics\cite{Lederman} predicts scaling as a function of $\displaystyle {t\over {t_w}}$.  Measurements of Ocio {\it et al.}\cite{Ocio,Vincent3}, exhibited in Figs. 3 and 4 of Sec. III show that this is not the case in the long time domain $t\geq {t_w}$.  That is, Fig. 4 corrects Fig. 3 for the stationary contribution.\cite{Bouchaud} Departure from scaling as $\displaystyle {t\over {t_w}}$ is only truly present for $t\geq {t_w}$, as can be seen in Fig. 4.  As is clearly seen in Fig. 4, the barrier hopping or the non-stationary dynamics results in the relaxation of older systems being ``faster'' when plotted versus $\displaystyle{t\over {t_w}}$ (although, of course, when plotted versus $t$, the older the system, the slower the relaxation).  Bouchaud {\it et al.}\cite{Hammann} recalled that all of the data of Alba {\it et al.}\cite{Alba,Ocio} could be rescaled with the response times in the spin glass scaling as $({t_w}+t)^\mu$, with $\mu<1$.  They noted that, from a more fundamental point of view, the scaling variable should be written as $\displaystyle{t\over {{{\tau^*}^{(1-\mu)}}{t_w^\mu}}}$, with $\tau^*$ a characteristic time scale.  Their manuscript ascribed a physical meaning to $\tau^*$, relating it to the {\it finite number of available metastable states} in real samples made of `grains' of finite size.  In their language, ``a finite size system will eventually find the `deepest trap' in its phase space, which corresponds to the equilibrium state.\cite{Bouchaud}  This will take a long, but finite time $t_{erg}$; when $t_w$ exceeds this `ergodic' time, $t_{erg}$, aging is `interrupted' because the phase space has been faithfully probed.  Beyond this time scale, conventional stationary dynamics resume.''

The precise origin of the ergodic time scale was ``...not easy to discuss since we do not know precisely what these `subsystems' are.''  Bouchaud {\it et al.}\cite{Hammann} suggested it could be magnetically disconnected regions (such as grains), the size of which was determined by sample preparation, and thus temperature independent; or it could be that the phase space of $3d$ spin-glasses is broken into mutually inaccessible regions (``true'' ergodicity breaking).  We argue below that finite size effects can indeed account for this behavior.  Our approach is the same as that of Bouchaud {\it et al.}\cite{{Bouchaud},{Hammann}}, but our arguments will be based on the barrier model of Sec. II.

The reasoning follows from the time dependence of the spin glass correlation length.  From the data in Fig. 2, and Eq. (2), $\xi({t_w},T)={(\displaystyle{{t_w}\over {\tau_0}})}^{0.153T/{T_g}}$, where $\tau_0$ is of the order of an exchange time ($\approx 10^{-12}~sec$).  This means that, for a given waiting time $t_w$, the correlation length $\xi({t_w},T)$ can be larger than a particle with size $r$.  All the available metastable states in that particle would be occupied in thermal equilibrium, and no further aging would take place.  The largest barrier in that particle has magnitude\cite{Wood2} $\Delta(r)=6.04{k_B}{T_g}\ell n\,r$, less than the largest barrier in the particles for which $\xi({t_w},T)<r$.  For these particles [large on the length scale of $\xi({t_w},T)$],
$\Delta({t_w},T)={k_B}T\ell\,n\Bigl({\displaystyle{{t_w}\over {\tau_0}}}\Bigr)$.
The characteristic time for decay of ${M_{TRM}}(t,{t_w},T)$ is proportional to $\Delta-{E_z}$.\cite{Kenning,Bouchaud}
Therefore, for small magnetic field changes, the characteristic decay time of the small particles, proportional to $\Delta(r)-{r^3}{\chi_{fc}}{H^2}$, is less than the decay time of the larger particles, proportional to $\Delta({t_w},T)-{\xi^3}({t_w},T){\chi_{fc}}{H^2}$.  Averaged over all particles, small and large, the characteristic decay time for ${M_{TRM}}(t,{t_w},T)$ will be less than $t_w$, the magnitude of the difference depending upon what fraction of the sample contains particles of size $r<\xi({t_w},T)$, {\it i.e.} the particle size distribution.  As $t_w$ increases, more of the particle sizes $r$ will be less than $\xi({t_w},T)$, thereby {\it shortening} the characteristic time for ${M_{TRM}}(t,{t_w},T)$ decay.  The characteristic time will shorten as $t_w$ increases, leading to a faster decay of $M(t,{t_w},T)$ with increasing $t_w$ when plotted as a function of $\displaystyle {t\over {t_w}}$.  This is exactly the effect posited by Bouchaud {\it et al.}\cite{Hammann}.  In addition, the Zeeman energy $E_z$ for the larger particles is proportional to ${\xi^3}({t_w},T)={N_s}$.  Increasing $t_w$ will increase $\xi({t_w},T)$, causing $E_z$ to increase with increasing $t_w$.  This further diminishes $\Delta({t_w},T)-{\xi^3}({t_w},T){\chi_{fc}}{H^2}$ for the larger particles, adding to the reduction of the characteristic time for decay of ${M_{TRM}}(t,{t_w},T)$ with increasing $t_w$.  This additional contribution has also been noted by Bouchaud {\it et al}.\cite{Hammann}\hfill\break
\par
\noindent
{\bf (c) Lack of scaling for ${\chi^{\prime\prime}}(\omega,t)$ at large $\omega t$}\hfill\break
\par
The jump in ${\chi^{\prime\prime}}(\omega,t)$ with the application of a magnetic field is an important test for any dynamical model.  Fig. 5 displays the {\it frequency} dependence of the change in ${\chi^{\prime\prime}}(\omega,t)$ when plotted against $\omega \cdot t$.  It is seen that the jump is larger, the smaller $\omega$.  The origin of this effect within the trap model\cite{Bouchaud} was associated with an increase of coupling to the magnetic field, the deeper the trap.
Within the barrier model,\cite{Joh} this effect was associated with an increase of coupling to the magnetic field, the higher the barrier.  The relationship of the jump in ${\chi^{\prime\prime}}(\omega,t)$ to these non-uniform magnetic field couplings arises naturally from the time dependence of the spin-glass correlation length [Eq. (2)], $\xi(t,T)$.  The smaller $\omega$, the larger $t$, when ${\chi^{\prime\prime}}(\omega,t)$ is plotted as a function of $\omega \cdot t$.  But larger $t$ means larger $\xi(t,T)$, thence larger ${N_s}~[~\propto {\xi^3}(t,T)]$, thence larger ${E_z}~(~\propto {N_s}{\chi_{fc}}{H^2})$.  This increase in $E_z$ with increasing $t$ maps directly onto the non-uniform magnetic field coupling of the trap model\cite{Bouchaud} and the barrier model.\cite{Joh}

There is, in addition to the non-uniform coupling to the magnetic field, an additional effect arising from the presence of crystallites with radius $r<\xi(t,T)$.  We shall show below that the magnitude of the jump in ${\chi^{\prime\prime}}(\omega,t)$ will also depend upon $\omega$ ({\it i.e.} violate scaling) when the spin glass correlation length becomes of the order of, or larger than, the size of a spin glass particle.  Conversely, the dependence of the change in ${\chi^{\prime\prime}}(\omega,t)$ upon $\omega$ over the full frequency regime could be used to generate the particle size distribution.

The effective waiting time, $t_w^{e\!f\!f}$, after a magnetic field change, is given by Eq. (1), allowing one to write,
$${t_w^{e\!f\!f}}={t_w}exp\Bigl(-{{E_z}\over {k_B}T}\Bigr)~~.\eqno(4)$$
This scaling was first established by Vincent {\it et al.}\cite{Bouchaud} and by Chu {\it et al.}\cite{Kenning}

This scaling can be incorporated into the {\it ac} susceptibility.  Before a {\it dc} magnetic field change, the relaxation of ${\chi^{\prime\prime}}(\omega,t)$ is well accounted for by a power law,\cite{Bouchaud}
$${\chi^{\prime\prime}}(\omega,t)-{\chi_{eq}^{\prime\prime}}\propto {(\omega t)^{-b}}~~,\eqno(5)$$
with $b>0$, and $\chi_{eq}^{\prime\prime}$ the equilbrium susceptibility ${\chi^{\prime\prime}}(\omega,t\rightarrow\infty)$.

Using the relationship Eq. (4), the change in ${\chi^{\prime\prime}}(\omega,t)$ upon a change in magnetic field at time $t_1$ can be written as,
$$\Delta {\chi^{\prime\prime}}(\omega,{t_1})={\chi^{\prime\prime}}(\omega,{t_1^{e\!f\!f}})-{\chi^{\prime\prime}}(\omega,{t_1})~~.\eqno(6)$$
Here, $t_1^{e\!f\!f}$ is the effective waiting time upon a magnetic field change, defined through Eq. (4) with ${t_w}\rightarrow{t_1}$.

With reference to Fig. 5, the lower the frequency $\omega$, the greater the time $t_1$ (because the abcissa is the scaling variable $\omega t$).  But the greater the time $t_1$ before the magnetic field is changed, the larger the spin glass correlation length $\xi({t_1},T)$, and therefore the more the likelihood that $\xi({t_1},T)>r$, the size of the spin glass particle.  But if $\xi({t_1},T)>r$, the effective response time, $t_1^{e\!f\!f}$, will be less than that for an ``infinite'' sample [$\xi({t_1},T)<r$], and ${\chi^{\prime\prime}}(\omega,{t_1^{e\!f\!f}})$ will be larger, increasing the size of the jump from Eq. (6).  However, ${\chi^{\prime\prime}}(\omega,{t_1})$ will also be slightly larger because it does not decay beyond ${\chi^{\prime\prime}}(\omega,{t_{erg}})$, where $t_{erg}$ is the time at which $\xi({t_1},T)=r$, decreasing the size of the jump from Eq. (6).

The increase in the first term in Eq. (6) turns out to be larger than the increase in the
second because finite size is involved in the argument of an exponential (through the Zeeman energy) in the first, while finite size enters only as an argument of a weak power law in the second (see the Appendix for details).  This results in a larger magnitude of the jump in ${\chi^{\prime\prime}}(\omega,{t_1})$, the larger $t_1$, or, equivalently, the smaller $\omega$, precisely what is seen experimentally in Fig. 5.

These arguments are qualitative, based upon Eqs. (4) and (5).  A quantitative evaluation of Eq. (6) is made in the Appendix, fully supporting the conclusions of this subsection.

Finally, for very short times, ${\chi^{\prime\prime}}(\omega,t)$ is seen to scale with $\omega\cdot t$ in Fig. 5.  Very short times means that $\xi(t,T)<r$ for all particles.  As we have already noted, under these conditions the barrier model calls for scaling with $\omega\cdot t$, again in agreement with experiment.

\section{Key Experiments and Numerical Simulations}

The arguments given above, especially in Sec. IV, are qualitative in nature, but fully capable of quantitative application.  There are two approaches which we feel would be most relevant.  The first, experimental, is one of careful magnetic field and waiting time variations upon a variety of samples.  The second would be to make use of the model of Sec. II to numerically simulate particular realizations of spin glass materials.\hfill\break
\par
\noindent
{\bf (a) Experimental determination of P(r)}\hfill\break
\par
Preliminary examination of a number of spin glass samples, using SEM techniques,\cite{Krassimir} suggest that many are made up of a powder-like array of small crystallites, embedding much larger apparently single crystal pieces.  Of course, this division may not be general, and may only apply to those materials with which we have been working.  Nevertheless, the analysis of Sec. IV, parts (a) and (b), can in this instance generate a measure of the bounds on the size distribution of the powder-like array component.

The analysis of Sec. IV, part (a) generates the upper end, $r_{max}$ of the small crystallite length scale distribution $P(r)$.  The experiments are at fixed waiting time, with measurements as a function of the change in magnetic field, $H$.
At very small magnetic field changes, only the volume occupied by ${\xi^3}({t_w},T)$ contributes to $E_z$, shifting the peak of $S(t)$ by reducing the energy difference $\Delta-{E_z}$.  As the magnetic field change increases, there will come a point when the energy difference $\Delta-{E_z}$ becomes comparable to $\Delta (r)-{{r_{max}}^3}{\chi_{fc}}{H^2}$ for the {\it largest} of the small crystallites.  We have argued in Sec. IV, part (a), that this causes the ``break'' in the plot of ${log_{10}}{t_w^{e\!f\!f}}~vs~H^2$.  Thus, the ``break'' field, $H_{break}$, generates a determination of $r_{max}$.

The other extreme of $P(r)$, $r_{min}$, can be extracted from the departure from scaling, exhibited in Figs. 3 and 4.  Here, the magnetic field change is fixed, and the experiments are a function of increasing waiting time, $t_w$.  From these two figures, the longer $t_w$ the more rapid the relaxation of ${M_{TRM}}(t,{t_w},T)$ as a function of the reduced time variable, ${\displaystyle {t\over {t_w}}}$.  For very small $t_w$, $\xi({t_w},T)$ is less than the ``minimum'' size of the powder-like array of small crystallites.  The departure from scaling occurs first when $t_w$ increases to a point where $\xi({t_w},T)={r_{min}}$, or equivalently $\Delta({t_w},T)=\Delta({r_{min}})$.  For longer $t_w$, the crystallite with dimension $r_{min}$ is at equilibrium, and for that part of the sample, aging is over.  Thus, the waiting time at which departure from scaling is first seen is a direct measure of $r_{min}$, the lower extreme of $P(r)$.

Careful (tedious!) measurements beginning from either domain, $r_{max}$ or $r_{min}$, can of course be used to generate all of $P(r)$.  At the very least, these two approaches will given a measure of the width of the $P(r)$ distribution.\hfill\break
\par
\noindent
{\bf (b) Numerical Simulations}\hfill\break
\par
An alternate, and certainly complementary approach, is to simulate the spin glass sample by selected choices for $P(r)$.    Previous simulations,\cite{Wood} using the barrier model of Sec. II, were able to duplicate the waiting time dependence of the response function $S(t)$.  Any attempt to fit to a particle size distribution $P(r)$ will require the ability to simulate ${log_{10}}{t_w^{e\!f\!f}}~vs~{H^2}$ and ${M_{TRM}}(t,{t_w},T)~vs~t$.  This procedure will not be unlike that of neutron or X-ray diffraction, where scattering from a specific model is measured against experiment.  It is the usual ``inverse'' problem where small iterations from a hypothesized model are used to fit experiment.  Previous success at fitting $S(t)$ suggests that a similar procedure, using the data of Sec. III and the analysis of Sec. IV, will be successful.  Having the limits $r_{max}$ and $r_{min}$ on $P(r)$ in hand, as discussed in part (a) of this Section, will greatly aid such an analysis.

\section{Conclusion}

This paper discusses spin glass dynamics for crystallites or amorphous particles of finite size.  Departures from scaling arise when the spin glass correlation length becomes of the order of or larger than particle sizes.  Qualitative arguments are given for the associated existence of a ``break field,'' $H_{break}$ away from a linear plot of ${log_{10}}{t_w^{e\!f\!f}}~vs~{H^2}$; the departure from $\displaystyle{t\over {t_w}}$ scaling of ${M_{TRM}}(t,{t_w},T)$; and the frequency dependence of the magnitude of the jump in ${\chi^{\prime\prime}}(\omega,t)~vs~\omega\cdot t$.  A guide to future experiments and numerical simulations, leading to the extraction of the particle size distribution, $P(r)$, are given with specific attention to what can be learned from experimental protocols.  SEM measurements\cite{Krassimir} of the particle size distribution will lead to explicit experimental consequences, setting the stage for a consistency check on the entire model.  The authors find it remarkable that the behavior of magnetization measurements in the time domain could so directly depend upon the physical size parameters of the sample particulates.

The authors have benefited from extensive discussions with Dr. J.-P. Bouchaud, experimental data supplied by Dr. M. Ocio, and from the financial support of the Japan Ministry of Education (Monbusho) and the U.S. National Science Foundation, Grant DMR 96 23195.

\par
\centerline{\bf Appendix}
\vskip .1cm
Sec. IV, subsection (c) gives a qualitative argument for the frequency dependence of the jump ${\chi^{\prime\prime}}(\omega,t)~vs~\omega t$ upon a change in magnetic field as a consequence of finite spin glass particle size.  This Appendix develops quantitative expressions for these quantities.

To derive the dependence of the jump in ${\chi^{\prime\prime}}(\omega,t)$ upon change in magnetic field on particle size, consider two cases: I. $r<\xi(t,T)$ and II. $r>\xi(t,T)$, where $r$ is the radius of the spin glass particle, and $\xi(t,T)$ the spin glass correlation length displayed in Eq. (2).  We shall assume that the time dependence of ${\chi^{\prime\prime}}$
is given by
$${\chi^{\prime\prime}}={\chi_{eq}^{\prime\prime}}(\omega)+A{(\omega t)^{-b}}~~,\eqno(A1)$$
where ${\chi_{eq}^{\prime\prime}}(\omega)$ may be different for cases I and II, but will cancel when we consider only the {\it change} in ${\chi^{\prime\prime}}(\omega,t)$ upon a change in magnetic field.  That is, we assume that the equilibrium ($t\rightarrow\infty$) value of ${\chi^{\prime\prime}}(\omega)$ is magnetic field independent.  The exponent $b\approx 0.20$ from experiment.\cite{Bouchaud}\hfill\break
\par
\noindent
I.  ${\rm particle~size}~r<\xi(t,T)$\hfill\break
\par
Consider the effect of a jump in magnetic field at a time $t_1$, and for this subsection, assume that the particle size $r<\xi({t_1},T)$.  Before the jump in magnetic field,
$${\chi_{I,before}^{\prime\prime}}(\omega\cdot{t_1})={\chi_{eq_I}^{\prime\prime}}(\omega)+A{(\omega\cdot {t_{erg}})^{-b}}~~,\eqno(A2)$$
where we have used $\xi({t_1},T)=({t_1}/{\tau_0})^a$, with $a=0.153T/{T_g}$ from Eq. (2), and where $t_{erg}={\tau_0}{r^{1/a}}$ is defined in Sec. IV, subsection (b).

The effective time, $t_I^{e\!f\!f}$ after a jump in magnetic field, is given by
$${t_I^{e\!f\!f}}={t_{erg}}exp{\Bigl(}-{{E_z}\over {k_B}T}\Bigr)~~.\eqno(A3)$$
This scaling was first established by Vincent {\it et al.}\cite{Bouchaud} and by Chu {\it et al.}, giving\cite{Kenning}
$${t_I^{e\!f\!f}}={\tau_0}{r^{1/a}}exp{\Bigl(-}{{{r^3}{\chi_{fc}}{H^2}}\over {{k_B}T}}{\Bigr)}~~.\eqno(A4)$$
Thus, after the jump in magnetic field,
$${\chi_{I,after}^{\prime\prime}}(\omega\cdot{t_1})={\chi_{eq_I}^{\prime\prime}}(\omega)+A{(\omega\cdot{t_I^{e\!f\!f}})^{-b}}~~.\eqno(A5)$$
Subtracting Eq. (A2) from Eq. (A5) gives the jump in ${\chi^{\prime\prime}}(\omega,{t_1})$ upon a jump in magnetic field:
$$\Delta{\chi_I^{\prime\prime}}(\omega,{t_1})=A{{\Bigl(}\omega{\tau_0}{r^{1/a}}{\Bigr)}^{-b}}{\Bigl[}exp\Bigl({b{r^3}{\chi_{fc}}
{H^2}\over {{k_B}T}}\Bigr)-1{\Bigr]}~~.\eqno(A6)$$
\par
\noindent
II.  $\xi(t,T)<{\rm particle~size}~r$\hfill\break
\par
\noindent
Consider the effect of a jump in magnetic field at a time $t_1$, and for this subsection, assume that $\xi({t_1},T)$ is smaller than the particle size.  Before the jump in magnetic field,
$${\chi_{II,before}^{\prime\prime}}(\omega\cdot{t_1})={\chi_{eq_{II}}^{\prime\prime}}(\omega)+A{(\omega\cdot {t_1})^{-b}}~~.\eqno(A7)$$
After the jump in magnetic field,
$${\chi_{II,after}^{\prime\prime}}(\omega\cdot{t_1})={\chi_{eq_{II}}^{\prime\prime}}(\omega)+A{(\omega\cdot{t_{II}^{e\!f\!f}})^{-b}}~~,\eqno(A8)$$
where,
$${t_{II}^{e\!f\!f}}={t_1}exp\Bigl(-{{\xi^3}{\chi_{fc}}{H^2}\over {k_B}T}\Bigr)~~.\eqno(A9)$$
The jump in ${\chi_{II}^{\prime\prime}}(\omega\cdot{t_1})$ is then given by subtracting Eq. (A7) from Eq. (A8):\hfill\break
$$\Delta{\chi_{II}^{\prime\prime}}(\omega\cdot{t_1})=A{\Bigl(\omega{\tau_0}{\alpha^{1/a}}{r^{1/a}}\Bigr)^{-b}}$$
$$\times \Bigl[exp\Bigl({b{\alpha^3}{r^3}{\chi_{fc}}{H^2}\over {k_B}T}\Bigr)-1\Bigr]~~.\eqno(A10)$$
For convenience, $\xi=\alpha r,~\alpha>1$; $\alpha$ is a function of ${t_1},T$; and ${t_1}={\tau_0}(\alpha r)^{1/a}$.

Experimentally, from Fig. 5, the magnitude of the jump in ${\chi^{\prime\prime}}(\omega,{t_1})$ is larger, the smaller $\omega$.  But the lower the frequency $\omega$, the greater the time $t_1$ (because the abscissa is the scaling variable $\omega t$).  And the greater the time $t_1$ before the magnetic field is changed, the larger the spin glass correlation length $\xi ({t_1},T)$, and therefore the more the likelihood that $\xi ({t_1},T)>r$, the size of the spin glass particle.

This means that the jump in ${\chi^{\prime\prime}}(\omega,{t_1})$ for case I should exceed the jump in ${\chi^{\prime\prime}}(\omega,{t_1})$ for case II, or more simply, that Eq. (A6) should exceed Eq. (A10), more for smaller $\omega$, or equivalently, larger $t_1$.

It is a somewhat tedious algebraic exercise, but one can show that this is indeed the case.  Thus, finite size effects can generate the $\omega$ dependence of the jump in ${\chi^{\prime\prime}}(\omega,{t_1})$.  This is a consequence of different $E_z$ values [through Eq. (A3)] as a consequence of differing particle sizes.  This feature\cite{Hammann} adds to the non-uniform magnetic field couplings (larger, the larger the trap depth) introduced in the trap model\cite{Bouchaud} or (larger, the larger the barrier height) introduced in the barrier model,\cite{Joh} independent of possible finite size effects, arising from the time dependence of $\xi(t,T)$ and hence of $E_z$.

\end{document}